\documentclass[         %
aps,                    
prd,                    
showpacs,               
superscriptaddress,    
nofootinbib,            
showkeys,               %
preprintnumbers,        %
floatfix]               
{revtex4}               
\usepackage{graphicx,longtable,amsmath}

\numberwithin{equation}{section}

\begin{document}
\title{Neutrino Capture on $^{13}$C}
\author{T. Suzuki}
\email{suzuki@phys.chs.nihon-u.ac.jp}
\affiliation{Department of Physics, College of Humanities and Sciences, Nihon University, 
Sakurajosui 3-25-40, Setagaya-ku, Tokyo 156-8550, Japan}
\affiliation{National Astronomical Observatory of Japan 2-21-1 
Osawa, Mitaka, Tokyo, 181-8588, Japan}
\author{A.~B. Balantekin}
\email{baha@physics.wisc.edu}
\affiliation{Department of Physics, University of Wisconsin, Madison, WI 53706, USA}
\affiliation{National Astronomical Observatory of Japan 2-21-1 
Osawa, Mitaka, Tokyo, 181-8588, Japan}
\author{T. Kajino}
\email{kajino@nao.ac.jp}
\affiliation{National Astronomical Observatory of Japan 2-21-1 
Osawa, Mitaka, Tokyo, 181-8588, Japan}
\affiliation{Department of Astronomy, School of Science, 
University of Tokyo, 7-3-1 Hongo, Bunkyo-ku, Tokyo, 113-0033, Japan}

\date{\today}
\begin{abstract}

We present neutrino cross sections on $^{13}$C. The charged-current cross sections leading to various  
states in the daughter $^{13}N$ and the neutral-current cross sections leading to various states in the daughter 
$^{13}$C are given. We also provide simple polynomial fits to those cross sections for quick estimates of the reaction  rates. We briefly discuss possible implications for the current and future scintillator-based experiments. 

\end{abstract}
\medskip
\preprint{}
\maketitle

\vskip 1.3cm

\section{Introduction}
\label{Section: Introduction}

Neutrino cross sections on $^{12}$C are extensively studied since many liquid scintillators and liquid track detectors proposed or currently used for neutrino experiments are based on various carbon compounds. The natural abundance of $^{13}$C is 1.07\%, hence a sizable detector would already contain a substantial amount of this isotope. As the neutrino physics moves to the precision stage, it is becoming more important to have a more precise knowledge of neutrino cross sections on $^{13}$C.  Motivated by a proposal to use a $^{13}$C-enriched target as a solar neutrino detector \cite{Arafune:1988hx}, this reaction was first investigated in the early 1990's \cite{Fukugita:1989wv} both in the shell model using Cohen-Kurath wave functions \cite{Cohen:1965qa} and using the effective operator method \cite{Arima:1988xa}. Ground-to-ground state transition and inclusive cross sections were also worked out in Refs. \cite{Pourkaviani:1991pb} 
and \cite{Mintz:2000xv}, respectively. 

Currently, three reactor experiments using scintillators, Daya Bay \cite{Guo:2007ug}, Double Chooz \cite{Ardellier:2006mn}, and RENO \cite{Ahn:2010vy}, have the primary goal of measuring the third neutrino mixing angle, $\theta_{13}$. There may be opportunities to do other physics with these detectors. For example to test the original neutrino-antineutrino oscillations of Pontecorvo 
\cite{Pontecorvo:1957cp}, i.e to look for the appearance of electron neutrinos in the reactor 
{\it antineutrino} flux, naturally occurring $^{13}$C in the scintillators may be an attractive target.  
Note that the threshold for both electron neutrino and electron antineutrino charged-current scattering on $^{12}$C is slightly more than  $\sim 13$ MeV, beyond the reach of reactor neutrinos. Similarly, electron antineutrino charged-current scattering threshold for $^{13}$C is also slightly more than  $\sim 13$ MeV, again beyond the reach of reactor neutrinos.
Hence $^{13}$C is an attractive target to detect very low-energy neutrinos. In this case, the only background is $^8$B solar neutrinos as geoneutrinos are all electron antineutrinos. 
Utility of scintillators to reconstruct supernova neutrino spectra has recently been discussed \cite{Dasgupta:2011wg}. 
If there is a supernova explosion during the operation of any scintillator-based experiment, $^{13}$C may again be useful to sort out fluxes of various flavors. There also are new proposals such as Low Energy Neutrino Observatory (LENA) \cite{Wurm:2011zn}  to build scintillator-based multipurpose neutrino observatories. 
To help these applications, we present an updated calculation of the neutrino cross sections on $^{13}$C. 

We present our method of calculations in the next section. Section III includes our results for partial cross sections given in different forms with an eye towards aiding designs of future experiments. We conclude the paper in Section IV with a few brief remarks.

\section{Method of Calculations}


A new shell-model Hamiltonian for $p$-shell nuclei, SFO\cite{Suzuki:2003},
is used to evaluate neutrino cross sections on $^{13}$C.
With this new Hamiltonian the magnetic properties of $p$-shell nuclei are considerably improved, 
for example, in calculating magnetic moments and Gamow-Teller transitions compared to
conventional Hamiltonians such as Cohen-Kurath (CK) Hamiltonian \cite{Cohen:1965qa}. 
Here, we study neutrino-nucleus reactions, which are induced mainly by 
excitations of Gamow-Teller and spin-dipole states. 

The SFO Hamiltonian, which was constructed for use in the $p$-$sd$ shell
configurations including up to 2-3 $\hbar\omega$ excitations, can
describe well the magnetic moments of $p$-shell nuclei systematically 
and Gamow-Teller (GT) transitions in $^{12}$C and $^{14}$C with a small
quenching for spin $g$-factor and axial-vector coupling constant: i.e., 
$g_{A}^{eff}$/$g_A$ = $g_s^{eff}$/$g_s$ =0.95. In SFO, the spin-isospin
components of the interaction, that is, the monopole term of the 
$0p_{1/2}$-$0p_{3/2}$ orbits in isospin $T$ =0 channel is enhanced 
as compared to the conventional interactions.  

Neutrino-nucleus reaction cross sections are evaluated using the 
multipole expansion of the weak hadronic currents,  
\begin{equation}
J_{\mu}^{C_{\mp}} = J_{\mu}^{V_{\mp}} + J_{\mu}^{A_{\mp}}
\end{equation}
for charge-exchange reactions ($\nu$, $\ell^{-}$) and ($\bar{\nu}, \ell^{+}$),
and
\begin{equation}
J_{\mu}^{N} = J_{\mu}^{A_3} + J_{\mu}^{V_3} -2 \mbox{sin}^{2}\theta_{W}
J_{\mu}^{\gamma}
\end{equation}
for neutral-current reactions, ($\nu$, $\nu$') and ($\bar{\nu}$,
$\bar{\nu}$'), where $J_{\mu}^{V}$ and $J_{\mu}^{A}$ are vector and 
axial-vector currents, respectively, and $J_{\mu}^{\gamma}$ is the 
electromagnetic current with $\theta_{W}$ being the Weinberg angle.

The reaction cross sections induced by $\nu$ or $\bar{\nu}$ are given 
as follows \cite{Walecka:1975}: 
\begin{eqnarray}
& &\left(\frac{d\sigma}{d\Omega}\right)_{\frac{\nu}{\bar{\nu}}} =
\frac{G^2\epsilon k}{4\pi^2}\frac{4\pi}{2J_i+1} 
\{\sum_{J=0}^{\infty} \{(1+\vec{\nu}\cdot\vec{\beta})\nonumber\\ 
& &\mid\langle J_f \parallel M_J \parallel J_i\rangle\mid^2 
+ [1-\hat{\nu}\cdot\vec{\beta}+2(\hat{\nu}\cdot\hat{q})(\hat{q}
\cdot\vec{\beta})]\nonumber\\
& &\mid\langle J_f\parallel L_J \parallel J_i\rangle
\mid^2
- \hat{q}\cdot(\hat{\nu}+\vec{\beta}) 2 Re \langle J_f \parallel
L_J \parallel J_i\rangle \nonumber\\
& &\langle J_f \parallel M_J \parallel J_i
\rangle^{\ast}\} 
+ \sum_{J=1}^{\infty} \{[1-(\hat{\nu}\cdot\hat{q})(\hat{q}\cdot
\vec{\beta})]\nonumber\\
& &(\mid\langle J_f \parallel T_{J}^{el} \parallel J_i
\rangle\mid^2 + \mid\langle J_f \parallel T_{J}^{mag} \parallel
J_i\rangle\mid^2 \nonumber\\
& &\pm \hat{q}\cdot(\hat{\nu}-\vec{\beta}) 2 Re [\langle J_f \parallel
T_{J}^{mag} \parallel J_i \rangle \nonumber\\ 
& &\langle J_f \parallel T_{J}^{el}
\parallel J_i \rangle^{\ast}])\} 
\label{eqq}
\end{eqnarray}
where $\vec{\nu}$ and $\vec{k}$ are the neutrino and lepton momenta,
respectively. In Eq. (\ref{eqq}),  $\epsilon$ is the lepton energy, $\vec{q}=\vec{k}
-\vec{\nu}$, $\vec{\beta}=\vec{k}/\epsilon$, $\hat{\nu}=\vec{\nu}
/\mid\vec{\nu}\mid$, and $\hat{q}=\vec{q}/\mid\vec{q}\mid$. 
For the charged-current reactions 
$G=G_{F}$cos$\theta_C$ with $G_{F}$ being the Fermi coupling constant, and  
$\theta_C$ is the Cabibbo angle. 
In this case, the label lepton refers to either the electron or the positron. 
For neutral current reactions, $G=G_{F}$  
and the lepton is scattered neutrino. To describe the final-state interactions, cross sections are multiplied by 
the Fermi function in case of the charge-exchange reactions.  
In Eq. (\ref{eqq}), $M_{J}, L_{J}, T_{J}^{el}$ and $T_{J}^{mag}$ are
Coulomb, longitudinal, transverse electric and magnetic multipole
operators for the vector and axial-vector currents. 

\begin{figure}[t]
\begin{center}
\includegraphics[scale=2.3]{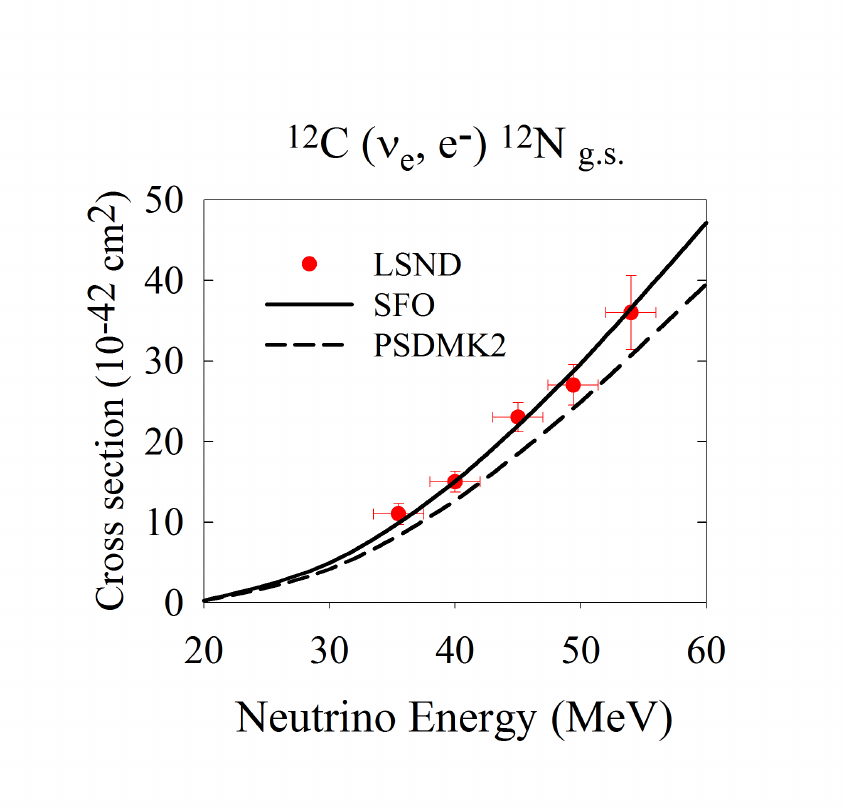}
\caption{Cross sections for the exclusive reaction, $^{1 2}$C ($\nu_e$, $e^{-}$) 
$^{12}$N (1$^{+}_{g.s.}$), obtained by shell-model calculations with the use
of SFO and PSDMK2. Experimental data are taken from ref. \cite{LSND:1997}. 
\label{fig:fig1}}
\end{center}
\end{figure}

The improvement achieved by the new Hamiltonian, SFO, is demonstrated in
Figure \ref{fig:fig1}, where calculated cross sections for the exclusive neutrino-induced
reaction $^{12}$C ($\nu$, $e^{-}$) $^{12}$N (1$^{+}_{g.s.}$) for the SFO and
CK type\footnote{In this particular case we use the so-called PSDMK2 Hamiltonian \cite{MK:1975,OXB:1986}.} Hamiltonians are shown.
Since the GT strength in $^{12}$C is well described by the SFO Hamiltonian \cite{Suzuki:2006qd}, the cross
sections obtained using the SFO Hamiltonian naturally reproduce the experimental 
data \cite{LSND:1997} very well. To explore this feature of the SFO Hamiltonian,
the cross sections calculated using it were applied to the calculation of the abundances resulting from 
neutrino-process nucleosynthesis in core-collapse supernovae in order to discuss the
effects of neutrino oscillation on the light element synthesis \cite{Yoshida:2008zb}. 

\begin{table}[t]
\begin{tabular}{ c | c | c | c } \hline
$B(GT$: $^{13}$C $\rightarrow$ $^{13}$N) & SFO & CK\cite{Fukugita:1989wv} 
& EXP. \\\hline  
$^{13}$N  $\hspace*{0.2cm}$ $J^{\pi}$ $\quad$ $E_x$ (MeV) &  &  &  \\
  1/2$^{-}$ $\hspace*{0.2cm}$ 0.0  & 0.284 & 0.420 & 0.411$\pm$0.004\cite{Ajzenberg:1986}\\
                     &       &       & 0.398$\pm$0.008\cite{Tadeucci:1987} \\
  1/2$^{-}$ $\hspace*{0.2cm}$ 8.92  & 0.569 & 0.524 & \\
  3/2$^{-}$ $\hspace*{0.2cm}$ 3.50   & 2.103 & 2.14 & 1.64$\pm$0.10\cite{Tadeucci:1987} \\
  3/2$^{-}$ $\hspace*{0.2cm}$ 9.46   & 0.500 & 0.260 & \\\hline
$B(M1)$ ($\mu_N^2$)&  & & \\
 $^{13}$C (3/2$^{-}$: 3.68 MeV) $\rightarrow$ $^{13}$C (1/2$^{-}_{g.s.}$) & 
0.878 & 1.17\cite{Arafune:1988hx} & 0.698$\pm$0.072 \\\hline 
\end{tabular}
\caption{
$B(GT)$ and $B(M1)$ values in $^{13}$C obtained by shell-model
calculations with the SFO and CK \cite{Fukugita:1989wv} Hamiltonians as well as the experimental ones.
} 
\label{tab:table1}
\end{table}

Calculated $B(GT)$ and $B(M1)$ values in $^{13}$C are given in Table I.
The values with the SFO Hamiltonian are obtained in the $p$-$sd$ shell including up to 
2$\hbar\omega$ excitations with $g_{A}^{eff}$/$g_A$ =0.95, while those
for the CK Hamiltonian are obtained within the $p$-shell with a larger quenching factor,
$g_{A}^{eff}$/$g_A$ =0.69\cite{Fukugita:1989wv}, which is adjusted to reproduce 
the experimental $B(GT)$ values of $^{13}$N($\beta^{+}$)$^{13}$C, 
$^{15}$O($\beta^{+}$)$^{15}$N and $^{11}$C($\beta^{+}$)$^{11}$B.
The SFO Hamiltonian can explain the experimental GT strength
for the transition to $^{13}$N (3/2$^{-}$, 3.50 MeV) rather well.
This transition is of interest as it may be possible to measure the $\gamma$ decay from $^{13}$N.  
The $B(M1)$ value for $^{13}$C (3/2$^{-}$, 3.68 MeV) $\rightarrow$ $^{13}$C 
(1/2$^{-}_{g.s.}$) obtained by the SFO Hamiltonian is also found to be close to the
experimental value. Here, the isovector spin $g$ factor is 
$g_s^{IV, eff}$/$g_s^{IV}$ =0.95 and $\delta g_{\ell}^{IV}$ is taken to 
be -0.15.        


\section{Results}

\begin{figure}[b]
\begin{center}
\includegraphics[scale=0.50]{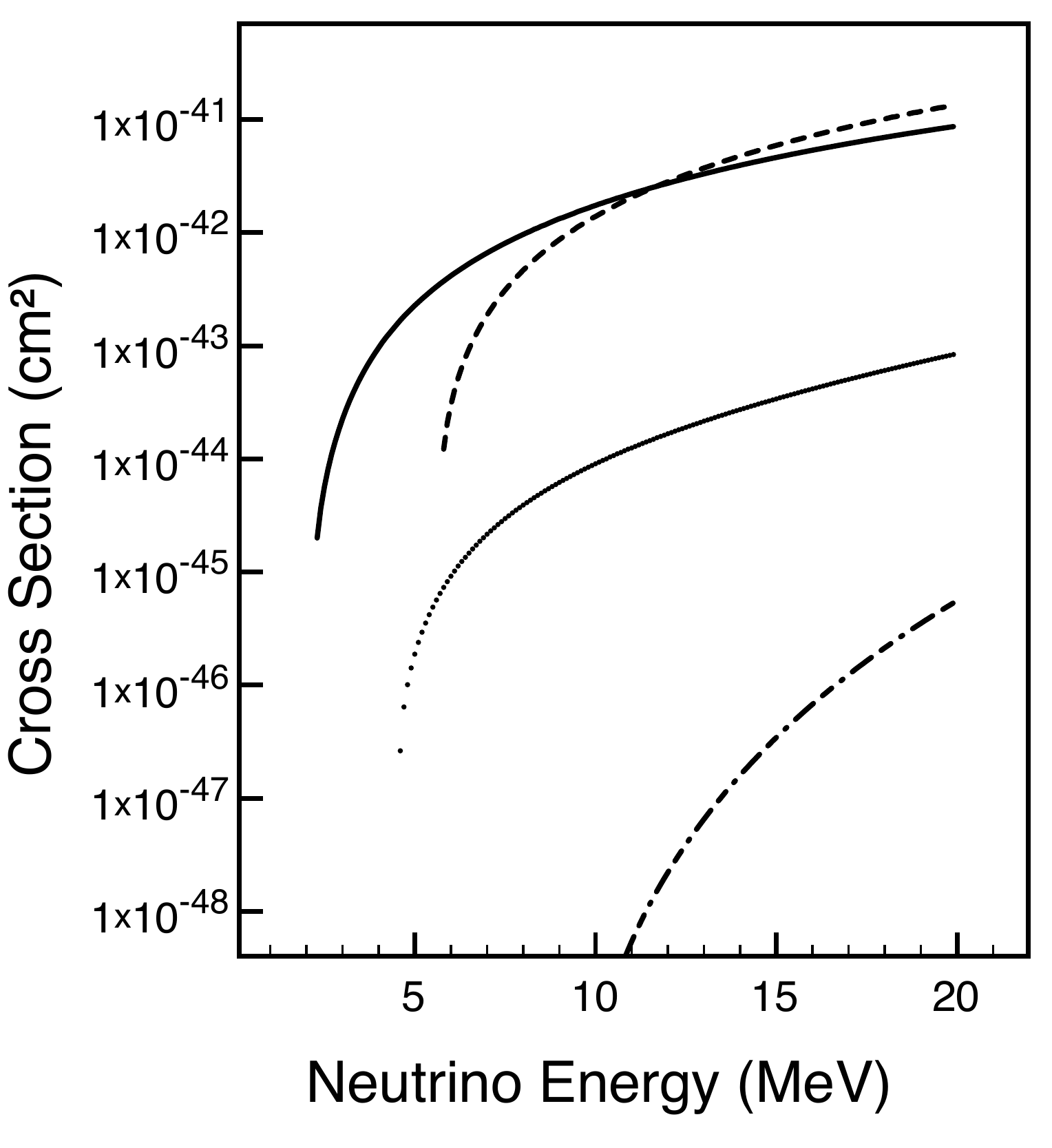}
\caption{Partial cross sections for the reaction, $^{13}$C ($\nu_e$, $e^{-}$) 
$^{13}$N, leading to various excited states in the daughter nucleus. The solid line is for the ground state. The dotted, dashed and dot-dashed lines are for the 2.365 MeV, 3.502 MeV, and 7.376 MeV excited states in $^{13}$N, respectively.} 
\label{fig:fig2}
\end{center}
\end{figure}

\begin{figure}[b]
\begin{center}
\includegraphics[scale=0.90]{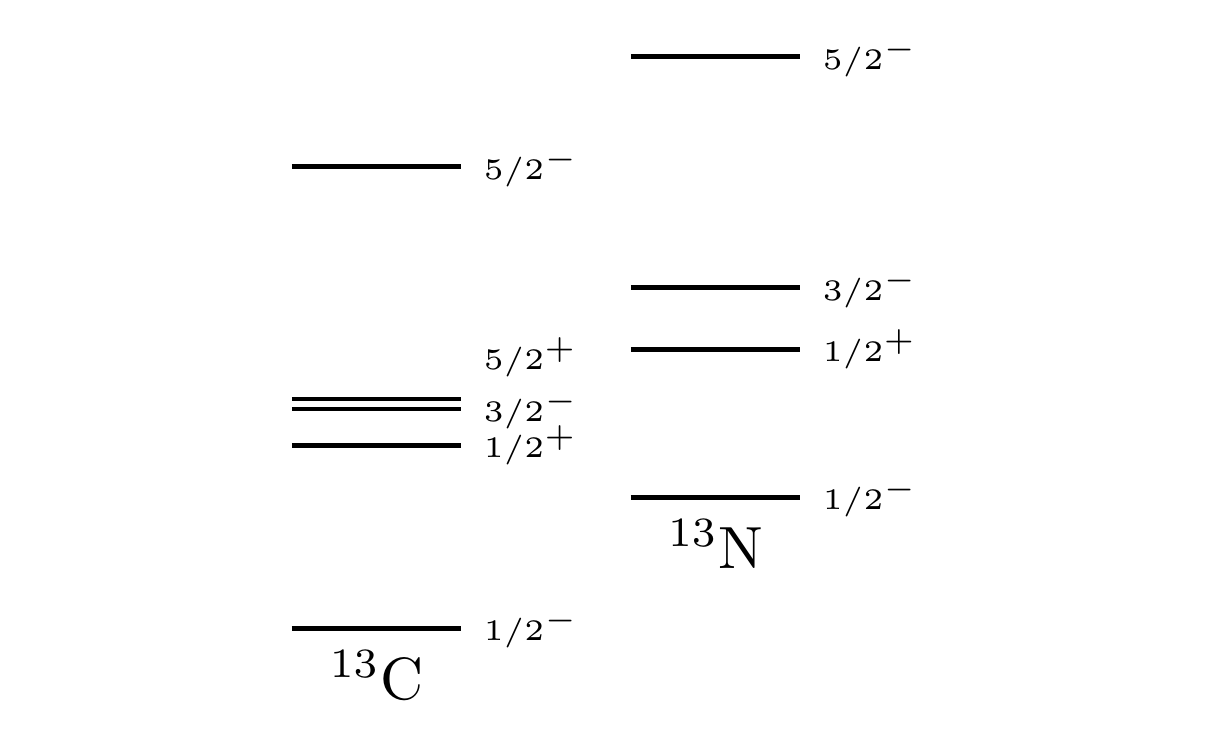}
\caption{States of the $^{13}$C and  
$^{13}$N nuclei included in this work.}
\label{fig:fig1a}
\end{center}
\end{figure}

We present the charged-current neutrino reaction cross sections on $^{13}$C target leading to a particular state in $^{13}$N in three different formats. Figure \ref{fig:fig2} displays these cross sections graphically for reactions leading to the ground state (1/2$^-$, solid line), as well as to the 1/2$^+$ ($E_x$= 2.365 MeV, dotted line),  
3/2$^-$ ($E_x$=3.502 MeV, dashed line), and the 5/2$^-$ ($E_x$=7.376 MeV, dot-dashed line) states in $^{13}$N. 
(Energy levels the $^{13}$C and  $^{13}$N nuclei included in this work are shown in Figure \ref{fig:fig1a}). 
Clearly the dominant contributions come from the transitions to the ground state and to the 3.502 MeV excited state in the daughter nucleus. We present those two cross sections in tabular form in Tables \ref{tablo1} 
and \ref{tablo2}, respectively. 

\begin{table}[t]
  \centering
  \begin{tabular}{@{} cccccc @{}}
    \hline
    $E_{\nu}$(MeV) & $\sigma$ (cm$^2$) &$E_{\nu}$(MeV) & $\sigma$ (cm$^2$)& $E_{\nu}$(MeV) & $\sigma$ (cm$^2$)
\\ 
    \hline
2.3& $2.00 \times 10^{-45}$ & 7.0 & $6.62 \times 10^{-43}$ & 12.0 & $2.73 \times 10^{-42}$\\
2.5& $5.66 \times 10^{-45}$ & 7.5& $8.07 \times 10^{-43}$ & 12.5&$ 3.02 \times 10^{-42}$ \\
    3.0& $2.25 \times 10^{-44}$ & 8.0 & $ 9.65 \times 10^{-43}$ &13.0 &$3.31 \times 10^{-42}$  \\ 
    3.5 & $5.23 \times 10^{-44}$  & 8.5 & $1.14 \times 10^{-42}$ & 13.5& $3.62 \times 10^{-42}$ \\ 
    4.0 & $9.60 \times 10^{-44}$& 9.0& $1.33 \times 10^{-42}$ & 14.0 & $3.94 \times 10^{-42}$\\ 
    4.5 & $1.54 \times 10^{-43}$& 9.5 & $1.53 \times 10^{-42}$ & 14.5 &$4.27 \times 10^{-42}$\\
5.0& $2.27 \times 10^{-43}$ & 10.0& $1.74 \times 10^{-42}$ & 15.0 & $4.62 \times 10^{-42}$\\
5.5&$3.14 \times 10^{-43}$ & 10.5&$1.97 \times 10^{-42}$& 15.5 & $ 4.98 \times 10^{-42}$\\
6.0&$ 4.16 \times 10^{-43}$& 11.0 &$ 2.21\times 10^{-42}$& 16.0 & $ 5.35 \times 10^{-42}$\\
6.5& $5.32 \times 10^{-43}$& 11.5 &$2.47 \times 10^{-42}$ &16.5& $ 5.73 \times 10^{-42}$\\
    \hline
  \end{tabular}
  \caption{Cross sections for the charged current reaction $^{13}$C ($\nu_e, e^-$) $^{13}$N  leading to the 1/2$^-$ (ground) state (with excitation energy 0.0 MeV). 
}
  \label{tablo1}
\end{table}

\begin{table}[htbp]
  \centering
  \begin{tabular}{@{} cccccc @{}}
    \hline
    $E_{\nu}$(MeV) & $\sigma$ (cm$^2$) &$E_{\nu}$(MeV) & $\sigma$ (cm$^2$)& $E_{\nu}$(MeV) & $\sigma$ (cm$^2$)
\\ 
    \hline
5.8& $1.22 \times 10^{-44}$ & 9.5 & $1.12 \times 10^{-42}$ & 13.5 & $4.22 \times 10^{-42}$\\
6.0& $3.01\times 10^{-44}$ & 10.0& $1.39 \times 10^{-42}$ & 14.0 &$ 4.76 \times 10^{-42}$ \\
  6.5& $9.35 \times 10^{-44}$ & 10.5 & $  1.70 \times 10^{-42}$ &14.5  &$5.32 \times 10^{-42}$  \\ 
    7.0 & $1.87 \times 10^{-43}$  & 11.0 & $2.05 \times 10^{-42}$ & 15.0 & $5.91 \times 10^{-42}$ \\ 
    7.5 & $3.11 \times 10^{-43}$& 11.5& $ 2.42 \times 10^{-42}$ & 15.5 & $ 6.54 \times 10^{-42}$\\ 
    8.0 & $4.66 \times 10^{-43}$& 12.0 & $2.82 \times 10^{-42}$ & 16.0 &$7.19 \times 10^{-42}$\\
8.5 & $6.51 \times 10^{-43}$ & 12.5& $  \times 3.2610^{-42}$ & 16.5 & $7.88 \times 10^{-42}$\\
9.0&$8.68 \times 10^{-43}$ & 13.0&$3.73 \times 10^{-42}$& 17.0 & $ 8.60 \times 10^{-42}$\\
    \hline
  \end{tabular}
  \caption{Cross sections for the charged current reaction $^{13}$C ($\nu_e, e^-$) $^{13}$N  leading to the 3/2$^-$ (ground) state (with excitation energy 3.502 MeV). 
}
  \label{tablo2}
\end{table}

Sometimes it is convenient to have an analytical form for the cross sections to do quick estimates for counting rates. Although  the cross sections grow like $\sim E^2$ at higher energies, because of the energy-dependence of the nuclear matrix elements, the energy-dependence of the cross sections near the reaction threshold energies is more  intricate. We find that the power series expansion 
\begin{equation}
\label{1}
\sigma = 10^{-44} {\rm cm}^2 [a_1 (E_{\nu}-Q) + a_2 (E_{\nu}-Q)^2 + a_3 (E_{\nu}-Q)^3 ], 
\end{equation}
where $Q$ is the Q-value of the reaction in MeV, is sufficiently accurate within few percent. In Table \ref{table:2} we present the values of the coefficients in Eq. (\ref{1}) for the charged-current cross sections. 

\begin{table}[t]
  \centering
  \begin{tabular}{@{} cccccc @{}}
    \hline
    $^{13}$N state & $E_x$ (MeV) & Q (MeV) & $a_1$ (MeV$^{-1}$) & $a_2$ (MeV$^{-2}$) & 
    $a_3$ (MeV$^{-3}$) \\
    \hline
    ${1/2}^-$ &0 & 2.2 & 0.16 & 2.86 & 0.00  \\
    ${1/2}^+$ & 2.365  & 4.59 & $6.50 \times 10^{-2}$ & $1.20 \times 10^{-2}$  & $1.26 \times 10^{-3}$   \\
    ${3/2}^-$ & 3.502 & 5.72 & 6.20 & 6.13 & 0 \\
    ${5/2}^-$ & 7.376 & 9.60 &  $1.59 \times 10^{-3}$ & $- 7.68 \times 10^{-4}$ & $1.07 \times 10^{-3}$ \\ 
    \hline
  \end{tabular}
  \caption{Parameter values in Eq. (\ref{1})  for the cross sections of the charged-current reaction $^{13}$C ($\nu_e, e^-$) $^{13}$N  going to the indicated state in $^{13}$N.}
  \label{table:2}
\end{table}

\begin{figure}[b]
\begin{center}
\includegraphics[scale=0.45]{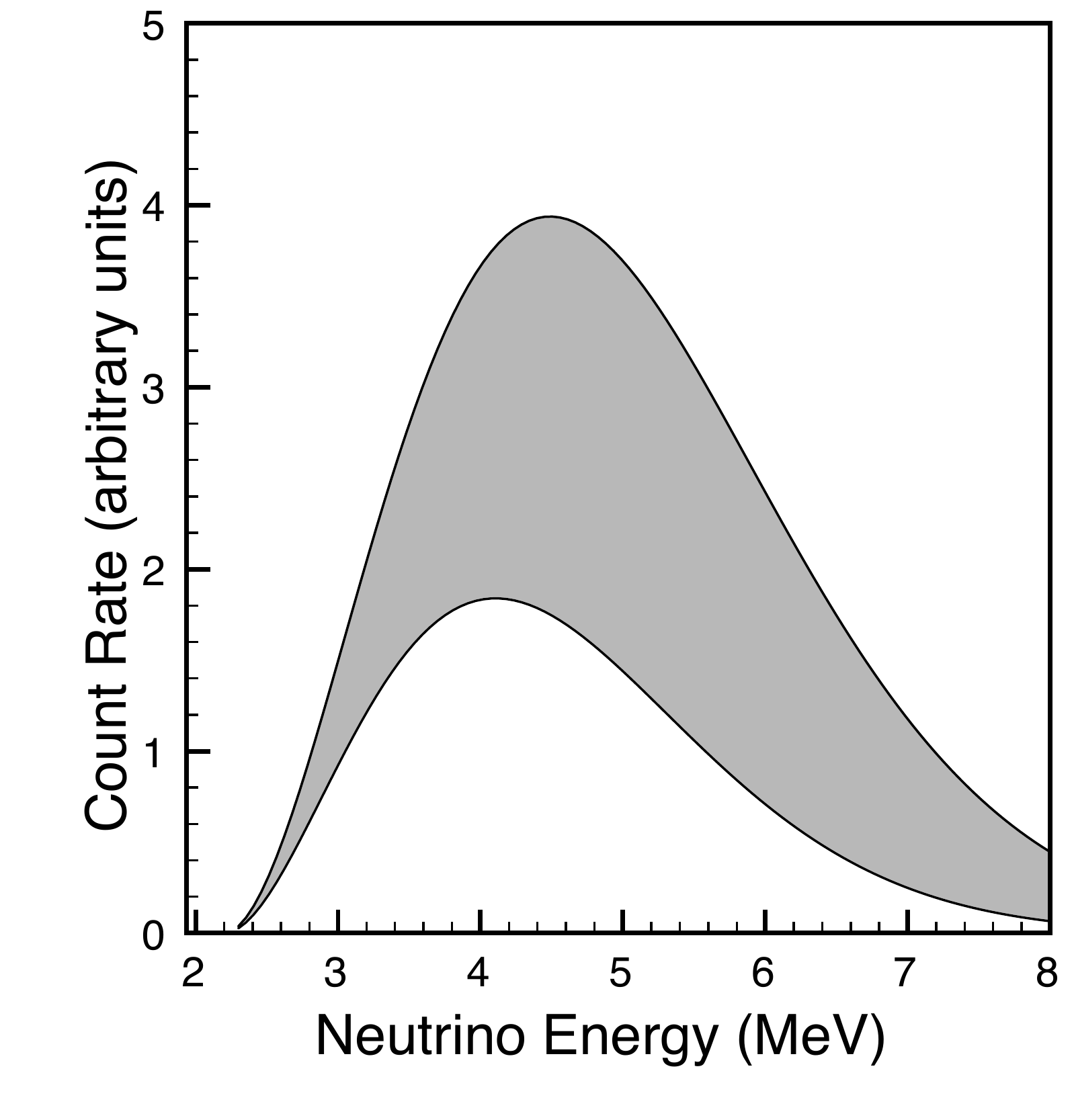}
\caption{An illustration of the spread of the electron neutrino count rate with different reactor neutrino flux models for the charged-current transition to the $^{13}$N ground-state. Models of the Refs. \cite{Davis:1979gg}, \cite{Vogel:1989iv}, and \cite{Huber:2011wv} are used.}
\label{fig:spread}
\end{center}
\end{figure}

One possible application of these cross sections could be to search for electron neutrino appearance in the reactor antineutrino flux. Ideally one would like to be as close to the reactors as possible much like experiments searching for the neutrino magnetic moment. However, in this case reactor neutrino flux uncertainties can be sizable. 
In Figure \ref{fig:spread}  we show the spread of the count rate for the charged-current transition to the $^{13}$N ground-state due to variations in various reactor neutrino spectrum models as a function of the neutrino energy. (We use the models of the Refs. \cite{Davis:1979gg}, \cite{Vogel:1989iv}, and \cite{Huber:2011wv} to illustrate the spread). We assume an ideal experiment with perfect energy resolution. The actual count rate will depend on the number of electron neutrinos present, but clearly there is a factor of two variation in the rate due to the uncertainties of the flux. (This is also true for the energy-integrated count rate, which is between 5.20 and 13.02 in the same arbitrary units). 
One possible way to reduce such uncertainties is to use much more abundant electron antineutrinos to estimate the reactor flux. This can be achieved using neutral-current scattering. Hence we also provide the neutral current cross sections below. 

\begin{figure}[t]
\begin{center}
\includegraphics[scale=0.50]{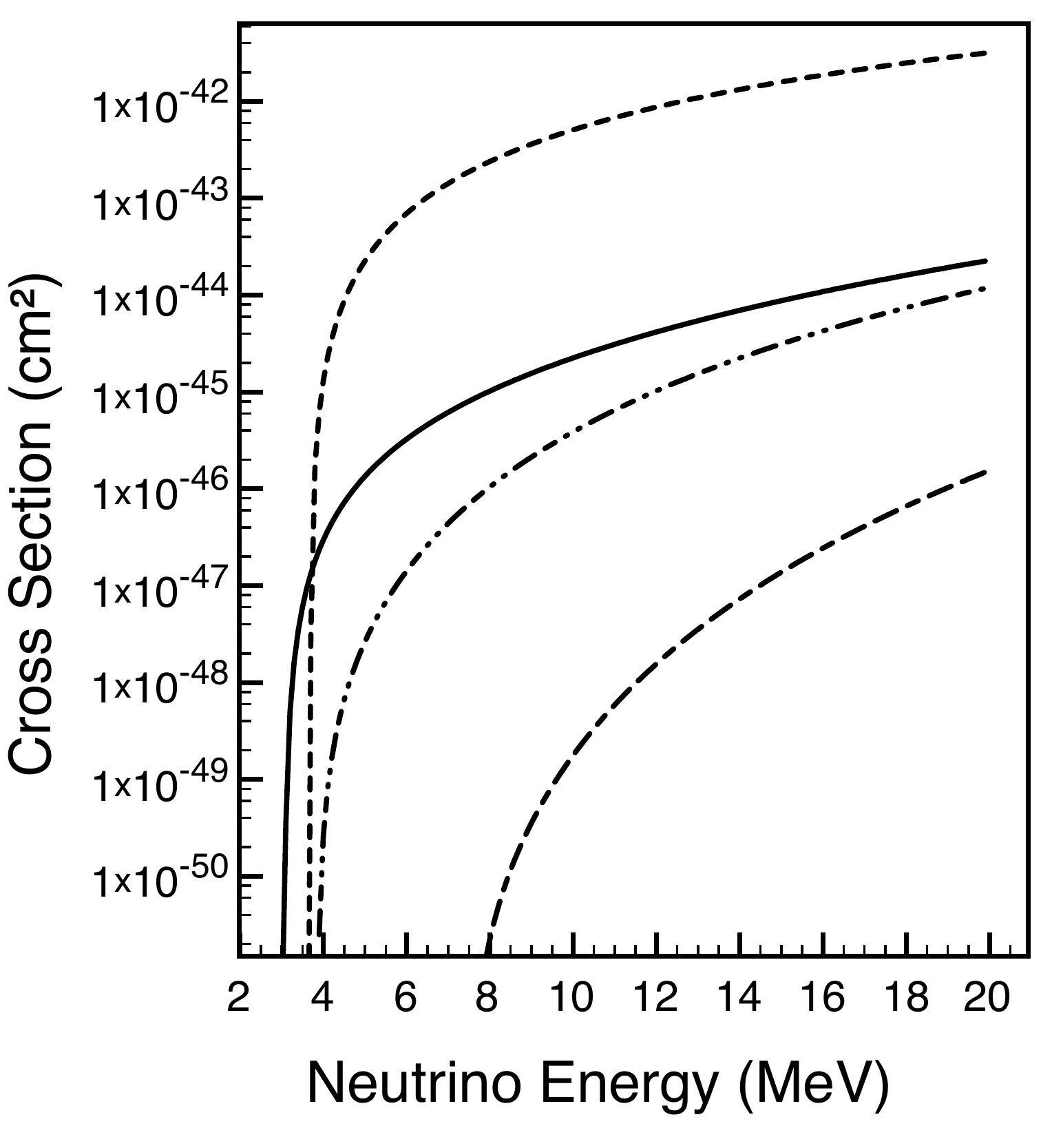}
\caption{Partial cross sections for the reaction, $^{13}$C ($\bar{\nu}_e$, $\bar{\nu}_e^{\prime}$) 
$^{13}$C, leading to various excited states in the daughter nucleus. The solid, dashed, dot-dashed (smallest cross section), and 
long-short dashed lines are for the excitation of 3.089 MeV, 3.685 MeV, 3.854 MeV and 7.547 MeV excited states in the $^{13}$C, respectively.} 
\label{fig:fig3}
\end{center}
\end{figure}

Electron antineutrino neutral-current cross sections for exciting various states in the daughter $^{13}$C are shown in Figure \ref{fig:fig3}\footnote{Note that cross sections for $(\nu_e, \nu_e^{\prime})$ are not much different from those for 
$(\bar{\nu}_e, \bar{\nu}_e^{\prime})$; they are larger only by about 2\%, 5\%, 8\% at 
$E_{\nu}$ = 5, 10 and 15 MeV, respectively.}. 
In this figure cross sections for the excitation of the 3.089 MeV 1/2$^+$ state (solid line), the 3.685 MeV 3/2$^{-}$ state (dashed line), the 3.857 MeV 5/2$^+$ state (dot-dashed line) and the 7.547 MeV 5/2$^-$ state (long-short dashed line) in $^{13}$C are shown. Clearly excitation of the  3.685 MeV 3/2$^{-}$ state dominates the neutral current scattering. We also fitted the neutral-current cross sections to the power series expansion in Eq. (\ref{1}). 
Q-values in this expression are simply the excitation energies in the daughter nucleus ($^{13}C$). The resulting parameter values are given in Table \ref{table:3}.

\begin{table}[b]
  \centering
  \begin{tabular}{@{} cccccc @{}}
    \hline
    State & $E_x$ (MeV) & $a_1$ (MeV$^{-1}$) & $a_2$ (MeV$^{-2}$) & 
    $a_3$ (MeV$^{-3}$) \\
    \hline
    ${1/2}^+$ & 3.089& $6.80 \times 10^{-3}$& $8.80 \times 10^{-4}$ & $4.00 \times 10^{-4}$  \\
    ${3/2}^-$ & 3.685 & 0.122 & 1.26 & 0   \\
    ${5/2}^+$ & 3.854 & $9.83 \times 10^{-3}$ & -$3.38 \times 10^{-3}$ & $4.54 \times 10^{-4}$ \\
    ${5/2}^-$ & 7.547 &  0.596 & -0.56 & 0.1 \\ 
    \hline
  \end{tabular}
  \caption{Parameter values in Eq. (\ref{1}) for the cross sections of the neutral-current reaction $^{13}$C ($\bar{\nu}_e$, 
  $\bar{\nu}_e^{\prime}$) $^{13}$C  going to the indicated state in the daughter nucleus.}
  \label{table:3}
\end{table}

\section{Conclusions}

In this article, we presented the charged-current neutrino cross sections on $^{13}$C leading to various  
states in the daughter $^{13}N$ and the neutral-current neutrino cross sections on the same nucleus leading to various states in the daughter $^{13}$C. To aid quick estimates of the reaction rates we also provided simple polynomial fits to those cross sections, accurate within three to four percent. A knowledge of these cross sections would help scintillator-based searches for low-energy electron neutrinos in environments dominated by the electron antineutrinos, such as nuclear reactors. In addition, it was pointed out that neutrino-proton elastic scattering in scintillator detectors would help measure total neutrino fluxes \cite{Beacom:2002hs}, especially for the neutrinos coming from a core-collapse supernova 
\cite{Dasgupta:2011wg}. Knowledge of the electron neutrino-$^{13}$C cross sections would also help to sort out the backgrounds for such a task. Although neutrino burst from neutralized supernova core-collapse 
provides a remarkable flux of electron neutrinos, the significant fraction of these 
neutrinos at the energies above ~14 MeV would induce both charged and 
neutral current interactions with $^{12}$C, masking a signal from $^{13}$C we 
discussed in this article.  It is also a challenge to experimentally separate the 
first few dozens of microsecond of electron neutrino burst from the flux of thermalized 
neutrinos of three flavors and their antiparticles.

\section*{Acknowledgments}

We thank K. Heeger, K. Kotake, K. Sumiyoshi, and T. Takiwaki for useful discussions. 
This work was supported in part by Grants-in-Aid for Scientific Research of the JSPS (200244035, 22540290) and for Scientific Research on Innovative Area of MEXT (20105004), in part
by the U.S. National Science Foundation Grant No. PHY-0855082, and 
in part by the University of Wisconsin Research Committee with funds
granted by the Wisconsin Alumni Research Foundation.


\end{document}